\documentstyle[]{article}
\title{ Intercommutation of Z-string loops violates baryon number }
\author{Jacek Dziarmaga  \\
        Jagellonian University, Institute of Physics,  \\
        Reymonta 4,30-059 Krak\'ow, Poland
        \thanks{E-mail address: ufjacekd@ztc386a.if.uj.edu.pl}\\}

\date{1 October 1994}
\begin{document}
\maketitle

    \begin{abstract}
    We show that delinking of Z-string loops changes helicity and thus violates
baryon number. The key point is that an unlinked vortex loop can not be
twisted.
Helicity of an eventual magnetic twist when averaged in time is zero.
    \end{abstract}

    Recently there has been wide interest in baryon number violating
processes in the standard electroweak theory. The first indication of such
nonperturbative processes was the discovery of the electroweak sphaleron
\cite{manton,manklink}. Another type of solution at first sight
completely different is the electroweak string \cite{nambu}.
The Z-string can also be interpreted as a two-dimensional
type of sphaleron \cite{olesen}. Weinberg-Salam model possesses all the
necessary ingredients to explain matter-antimatter asymmetry such as C and CP
violation and also there is a place for baryon number violating processes as
one can deduce from the anomaly equation
\begin{equation}\label{10}
  \partial_{\mu}j^{\mu}_{B}=\frac{N_{F}}{32\pi^{2}}
  (-g^{2}W^{a}_{\mu\nu}\tilde{W}^{a\mu\nu}
                             +g^{\prime 2}Y_{\mu\nu}\tilde{Y}^{\mu\nu}) \;,
\end{equation}
where
$\tilde{F}_{\mu\nu}=\frac{1}{2!}\varepsilon_{\mu\nu\rho\gamma}F^{\rho\gamma}$.
The R.H.S. of the equation can be rewritten as a total divergence and if
we assume that there is no baryon flux through boundaries we can relate
a change of the baryon number to a change of the Chern-Simons index of
the fields
\begin{equation}\label{20}
  \triangle Q_{B} = N_{F} \triangle(N_{CS}-n_{CS}) \;.
\end{equation}
The nonabelian $SU(2)_{L}$ Chern-Simons number is
\begin{equation}\label{30}
  N_{CS}=\frac{g^{2}}{32\pi^{2}}\int d^{3}x\; \varepsilon_{ijk}
   (W^{aij}W^{ak}-\frac{g}{3}\varepsilon_{abc}W^{ai}W^{bj}W^{ck}) \;,
\end{equation}
while its abelian $U(1)_{Y}$ counterpart reads
\begin{equation}\label{40}
  n_{CS}=\frac{g^{\prime2}}{32\pi^{2}}
                      \int d^{3}x\;\varepsilon_{ijk}(Y^{ij}Y^{k}) \;.
\end{equation}
The Chern-Simons numbers themselves are not gauge-invariant but once the gauge
is fixed there is a direct relation between the change of these indices
and the change of the baryon number (Eq. \ref{20}).

   In this note similarly as in \cite{vachaspati} and \cite{lo} we will
regard the bosonic part of W-S model as a massive classical background for
fermionic degrees of freedom. In the semiclassical framework we will restrict
to configurations with $W^{1}_{\mu}=W^{2}_{\mu}=0$. After orthogonal
transformation
\begin{eqnarray}\label{50}
  Z_{\mu}=\cos\theta_{W}W_{\mu}^{3}-\sin\theta_{W}Y_{\mu} \;,\nonumber\\
  A_{\mu}=\sin\theta_{W}W_{\mu}^{3}+\cos\theta_{W}Y_{\mu} \;
\end{eqnarray}
we will make further restriction to configurations with $A_{\mu}=0$ and
one-component Higgs field $\phi_{1}=0$, $\phi_{2}=\psi$. It is a
straightforward
calculation to check that if such restrictions are imposed as initial
conditions
on the fields and their time derivatives they are satisfied all through the
time evolution of the system.
Eq.(\ref{20}) now takes a simple form in terms of helicity $H_{Z}$
\begin{eqnarray}\label{60}
  \triangle Q_{B}=\frac{N_{F}}{16\pi^{2}}\alpha^{2}
                                \cos(2\theta_{W})\triangle H_{Z} \;,\nonumber\\
  H_{Z}=\int d^{3}x\; \vec{B}_{Z}\vec{Z} \;,
\end{eqnarray}
where $\alpha^{2}=g^{2}+g^{\prime 2}$.
In this way we have restricted to Abelian Higgs model
\begin{equation}\label{70}
  L=-\frac{1}{4}Z_{\mu\nu}Z^{\mu\nu}+D_{\mu}\psi^{\star}D^{\mu}\psi
                                               -V(\psi^{\star}\psi) \;\;,
\end{equation}
where $D_{\mu}\psi=\partial_{\mu}\psi+i\frac{\alpha}{2} Z_{\mu}\psi$ with
the coupling constant
$\alpha=g\cos\theta_{W}+g^{\prime}\sin\theta_{W}$. The potential
is
$V(\psi^{\star}\psi)=-\mu^{2}\psi^{\star}\psi+\lambda(\psi^{\star}\psi)^{2}$.

    By restriction to the Abelian Higgs model we are not able to say anything
conclusive about generic nonabelian configurations but our goal is
to show that at least within this framework helicity is not conserved during
delinking of string loops. This conclusion can not be obtained without careful
treatement of topology of the complex Higgs field. For a configuration
of the Higgs field to be well defined the phase of the field has to be
single-valued everywhere except the lines of vortices themselves. It is a
basic condition both for classical time-dependent solutions and for off-shell
configurations contributing to a path integral. To proceed further we need a
rather plausible dynamical assumption. Namely we assume that outside
of the finite-width vortex core the modulus of the Higgs field
approaches exponentially its vacuum expectation value with some characteristic
lenght which is small as compared to actual intervortex separations.
At the same time we assume that also the covariant derivative of the
Higgs field approaches exponentially zero outside of the core. The phase
$\omega$ and the gauge potential are related by
$Z_{\mu}=-\frac{2}{\alpha}\partial_{\mu}\omega$. In other words we assume
there is a finite-width core of a vortex and outside of the core
the energy density is approximately zero. These are characteristic
fictures of a straight Nielsen-Olesen vortex but also known exact or
approximate time-dependent solutions confirm this expectation
\cite{traw,silveira,waves,taubes}.

   Now we can look on vortex networks on a larger scale such that cores
are negligibly thin as compared to interstring distances. Discussion of core
effects is postponed until later. Under our assumption outside of the cores
the only relevant field variable is the phase $\omega$ while the gauge
potential is related to this phase by a "pure gauge" condition. As already
mentioned $\omega$ has to be single-valued everywhere except string lines.
At every moment of time we can find surfaces of a constant phase. The
surfaces can terminate only on string (defect) lines
or at infinity. For a $Z$-string along $z$-axis
\begin{equation}\label{80}
  \psi(r,\theta)=f(r)e^{i\theta} \;\;\;,\;\;\;
  Z_{\theta}=\frac{-a(r)}{\frac{\alpha}{2} r} \;\;,
\end{equation}
with boundary conditions $f(0)=0$, $f(\infty)=\sqrt{\frac{\mu^{2}}{2\lambda}}$
and $a(0)=0$, $a(\infty)=1$, the constant phase surfaces are semiplanes
of constant $\theta$ terminating on the string line and at spatial
infinity. One can perform a $U(1)$ gauge transformation on the fields
\begin{equation}\label{90}
  \psi^{\prime}=\psi e^{i\chi z} \;\;,\;\;
  Z^{\prime}_{\theta}=Z_{\theta} \;\;,\;\;
  Z^{\prime}_{3}=-\frac{\chi}{\frac{\alpha}{2}} \;\;.
\end{equation}
Now the surfaces are twisted. Helicity per unit lenght of such
a string is $\frac{8\pi\chi}{\alpha^{2}}$. The twist of the gauge field is
connected with a twist in the phase of the Higgs field.

   Now we can try to construct a p-fold twisted string loop. Let the string
configuration at time $t$ be $\vec{X}(t,\sigma)$. It is a vortex
so the phase winds around this line by $2\pi$. We chose a closed line
$\vec{X}^{\prime}(t,\sigma)$ close to the string line but not linked with
the string. What is more we
demand that on a strip spanned by these two lines the phase is single valued.
As such a line is followed in the direction of the twist the phase rises
by $2\pi p$. The above is just a definition of what we mean by a twisted string
loop. On the closed line $\vec{X}^{\prime}(t,\sigma)$ we can span a smooth
surface $S$. As the circulation of the phase along its edge is nontrivial for
nonzero twist $p$ there must exist at least $\mid p\mid$ points in its
interior where the phase is not single-valued. Circulation of the phase
is concentrated in these points. Points with k-fold circulation around them
are counted k-times. Now the surface can be slightly deformed
and once again we have at least $\mid p\mid$ singular points. By continuously
varing the surface we can construct lines of defects. They can be
admitted provided the moduli of the Higgs field vanishes on these lines.
In other words the lines are nothing else but vortices. Thus a p-fold twisted
vortex loop can exist but only if it is stuck on a bundle of $\mid p\mid$
vortices. In a world where the finiteness of energy condition admits only
vortex loops but not infinite strings the necessary
condition for a loop to be p-fold twisted is that it is linked with
at least $\mid p\mid$ other loops with the same orientation. A string loop
which is not linked with other loops can not be twisted.

   Now let us consider a special case of two loops linked once. Each of them
must be 1-fold twisted. The question is whether the twist of a phase along
a given string is necessarily related to helicity (Eq.(\ref{60})). The positive
answer is suggested by the minimal energy configuration in Eq.(\ref{90}).
It is indeed so as can be shown with first principles. A 1-fold twisted
string loop must be stuck on another vortex. The total magnetic flux through
a surface spanned on the loop $\vec{X}^{\prime}(t,\sigma)$ is equal to
$\stackrel{+}{-}\frac{4\pi}{\alpha}$ with the sign dependent on relative
orientation. Inside of the core of the vortex loop we can follow the lines of
magnetic field. As the magnetic field is confined to the core each of its lines
is stuck on the other vortex. For a choosen line magnetic field can
be approximated by
\begin{equation}\label{91}
  \delta\vec{B}=\delta\Phi\;\vec{t}\;\delta[\vec{x}-\vec{Y}(t,\sigma)] \;\;,
\end{equation}
where $\vec{t}$ is a unit vector tangent to the magnetic flux line $\vec{Y}$
and $\delta\Phi$ is a part of the total flux in a given line. The
contribution to helicity (\ref{60}) from the line is
\begin{equation}\label{92}
  \delta H= \delta B \int\;dl \vec{t}\vec{Z}=
   \delta B (\frac{4\pi}{\alpha}+\phi)
\end{equation}
$\phi$ is a total flux of magnetic lines being linked with a given line.
The first contribution is due to the vortex the loop is stuck on. When
contributions from all magnetic field lines are put together the total
helicity from a region around the loop due to the linking with the other loop
is $\stackrel{+}{-}(\frac{4\pi}{\alpha})^{2}$. Under the plausible
rectrictions discussed at the begining this part of helicity
relies merely on topology of the vortex network.

   If two string loops intercommute they form a single vortex loop.
Now the topological contribution to helicity is zero. Lo \cite{lo} uses
a magnetohydrodynamical (MHD) analogy to decide on how the magnetic field lines
are reconnected during intercommutation. His argument suggests that
the initial topological helicity is transformed into helicity associated
with a twist of magnetic field lines. Now we will show that helicity of such
a magnetic twist when averaged in time is zero.

  Small fluctuations around the background of the Nielsen-Olesen vortex
(\ref{80}) were analysed in \cite{arodz}. The deformations of the fields
were taken as
\begin{equation}\label{a10}
  A^{\alpha}=W^{\alpha}(t,z)u(x,y) \;\;,
\end{equation}
where Greek indices mean $t$ or $z$. Other field components are unchanged
to leading order. Field equations linearised in the above fluctuations are
\begin{eqnarray}\label{a20}
   \partial_{\alpha}W^{\alpha}=0 \;\;,\nonumber\\
  -\partial_{\beta}\partial^{\beta}W^{\alpha}=m^{2}W^{\alpha} \;\;,\nonumber\\
  -\triangle u+M^{2}_{Z}f^{2}(r)u=m^{2}u \;\;,
\end{eqnarray}
where $m^{2}$ is a separation constant and $\triangle$ is a Laplacian in $x,y$.
The second equation is the planar Schrodinger equation which is well known to
have at least one bound state with an eigenvalue
$0<m^{2}<M^{2}_{Z}$, which can be interpreted as a mass of the
gauge field trapped within the core. With such an excitation the total
helicity is
\begin{equation}\label{a30}
   H(t)=\int\;\;dxdydz\;(B^{3}Z^{3}+B^{\theta}Z^{\theta}) \;\;,
\end{equation}
where both $Z^{3}$ and $B^{\theta}$ come only from the excitation
while other components are those of the background. For a localised magnetic
twist \cite{davis} the integral along z-axis is convergent. With a use of the
second equation in the set (\ref{a20}) and one integration by parts one easily
obtains an equation of motion for the total helicity
\begin{equation}\label{a40}
   \frac{d^{2}}{dt^{2}}H(t)=-m^{2}H(t) \;\;.
\end{equation}
Helicity oscilates around zero $H(t)=H_{0}\cos(mt)$ with $H_{0}$
being its initial value. The time-dependence is nontrivial because of nonzero
$m^{2}$ - even the gauge field trapped within the vortex core has nonzero
mass. This result contradicts MHD analogy as such so it may be
also dubious if there is any nonzero $H_{0}$ right after intercommutation.
In MHD magnetic flux simply drifts together with the fluid and here it has
its own fully relativistic dynamics. Even if the initial value of $H$ is
nonzero its time average vanishes $<H>=0$. Thus the internal magnetic
helicity, even if oscillations are not dumped by some dissipation process,
is not related to the net baryon number. The net baryon number is related
to and only to the topological helicity or in other words linking of string
loops. Thus delinking of a pair of string loops by intercommutation changes the
average  helicity from its initial value given by topology to zero. The net
baryon number is violated in this process. Intercommutation of Z-string loops
violates baryon number.

   To analyse changes of the topological helicity it is convinient to
represent vortices as ribbons. One edge of a ribbon shold be identified
with a line of vanishing Higgs field. The rest of the ribbon should
coincide with a surface of constant phase $\omega=\pi$ say. A string
loop is p-fold twisted if the two edges of the ribbon are linked p-times.
An isolated and 1-fold twisted vortex ring would be represented by a Mobius
strip. It is not possible to span a surface of constant say zero phase on one
edge of Mobius strip without the other edge cutting it but if it cuts it
will not be any longer a surface of constant phase. This is a new formulation
of our previous argument that an unlinked vortex loop can not be twisted.

    The initially linked ribbons have to be twisted. The directions
of the twists have to be correlated with the orientation of the link.
One can construct a loop after intercommutation in a following way. Staple
the ribbons together in an antiparallel fasion - strings do rearrange in this
way just before intercommutation \cite{shellard}. The ribbons concide with
a constant phase thus the edges of zero Higgs field have to be put together.
Now cut the stapled part in the middle - real antiparallel string segments
annihilate \cite{shellard}. The result is a single untwisted ribbon as it
should be according to our discussion. The initial twists have undone
one-another during intercommutation.

      Turning this around one can take two separate ribbon loops
which have to be untwisted. One can staple them together in an
antiparallel fasion, cut and reconnect the ends. What one obtains is an
untwisted single string loop. In this case intercommutation does not change
the topological helicity. Conversely if strings colliding with relativistic
velocities just pass one through another local twists by $2\pi$ should
appear to be in agreement with our discussion. The strings must reconnect
in such a way that the surfaces of constant phase change continuously.
This explains usefulness of the Christmas ribbons toy model, which has been
originally applied to reconnections of magnetic field lines in
magnetohydrodynamics \cite{pfister}.

    Now we will consider an example which strongly suggests that there
is a family of solutions which continuously interpolates between two linked
loops and two separate vortex loops. With a passage from the initial
configuration to the final one helicity smoothly changes from its initial
value to zero. Let us consider Bogomol'nyi limit \cite{bogom} of the Abelian
Higgs model in dimensionless units
\begin{equation}\label{110}
  L=-\frac{1}{4}Z_{\mu\nu}Z^{\mu\nu}+\frac{1}{2}D_{\mu}\psi^{\star}D^{\mu}\psi
                       -\frac{1}{8}(\psi^{\star}\psi-1)^{2} \;\;,
\end{equation}
with $D_{\mu}=\partial_{\mu}+iZ_{\mu}$. The model admits static planar
two-vortex solutions
\begin{equation}\label{120}
  \psi=\psi(x,y,\lambda_{A}]\;\;,\;\;
  Z_{\alpha}=Z_{\alpha}(x,y,\lambda_{A}]\;\;,\;\;
  Z_{k}=0 \;\;,
\end{equation}
where $\alpha,\beta ...$ mean $1,2$ while $k,l ...$ take values $0,3$.
$\lambda$'s are a set of four real parameters defining positions of vortices.
It was shown in \cite{waves} that for a coincident 2-vortex configuration
there are exact splitting modes in a form of travelling waves. The nonvanishing
fields in Eq.(\ref{120}) are modified by introducing time dependence through
the
parameters $\lambda_{A}=\lambda_{A}(t-z)$. The $0$ and $3$ components of gauge
potential take a form
\begin{equation}\label{130}
  Z_{k}=\sum_{A} F_{A}(x,y,\lambda_{A}]\partial_{k}\lambda_{A} \;\;,
\end{equation}
where the profile functions satisfy
\begin{equation}\label{140}
  (\frac{\partial}{\partial x^{\alpha}}\frac{\partial}{\partial x^{\alpha}}
   -\psi^{\star}\psi)F_{A}=
   -(\psi^{\star}\psi)\frac{\partial\omega}{\partial\lambda_{A}} \;\;,
\end{equation}
with $\omega$ being an actual phase.

    Let us consider the following vortex configuration
\begin{equation}\label{150}
  \psi=z(z+X) W(z,z^{\star}) \;\;\;,\;\;\;
  X=X^{1}+iX^{2} \;\;,
\end{equation}
where $W(z,z^{\star})$ is a real positive function and travelling
wave defined by $X=R\exp ik_{i}x^{i}$ with
$k_{i}k^{i}\equiv k_{0}^{2}-k_{z}^{2}=0$.
Such a solution is a vortex helix of radius $R$ winding around
a straight linear vortex and moving up the $z$-axis with a speed of light.

    Let us consider first a situation of large $R$ as compared to lenght
scales of the model. An approximate solution to Eq.(\ref{140}) is \cite{waves}
$F_{(X^{\beta})}=Z_{\beta}(\vec{x}+\vec{X})$ with $Z_{\beta}(\vec{x})$
being the potential of the single vortex solution (\ref{80}).
The $z$-component contribution to helicity per unit lenght is
\begin{equation}\label{160}
  \int d^{2}x\; Z^{3}B^{3}=
  -\int d^{2}x\; [B(\mid\vec{x}\mid)+B(\mid\vec{x}+\vec{X}\mid)]
                 \sum_{\beta} Z_{\beta}(\vec{x}+\vec{X})
                              \frac{\partial X^{\beta}}{\partial z} \;\;,
\end{equation}
Where $B(\mid\vec{x}\mid)$ is a magnetic field of the single vortex (\ref{90}).
Due to symmetries of the fields a contribution to the above integral from
around $-\vec{X}$ vanishes. For large $R$ we can approximate under the integral
$B(\mid\vec{x}\mid)\approx -2\pi\delta^{(2)}(\vec{x})$ and then perform the
integral with a result $2\pi\omega$, where $\omega=k_{0}$. Induced components
of magnetic field are $B^{\alpha}=-\sum_{\beta\gamma}
\varepsilon_{\alpha\beta}Z_{\beta\gamma}\frac{\partial X^{\gamma}}{\partial
z}$.
Their contribution to helicity is
\begin{equation}\label{170}
  \sum_{\alpha}\int d^{2}x\; Z^{\alpha}B^{\alpha}=
  \sum_{\alpha\beta\gamma}\int d^{2}x\;

[Z_{\alpha}(\vec{x})+Z_{\alpha}(\vec{x}+\vec{X})]\varepsilon_{\alpha\beta}
       Z_{\beta\gamma}(\vec{x}+\vec{X})\frac{\partial X^{\gamma}}{\partial z}
\;\;.
\end{equation}
Once again we can use symmetry properties of the fields and then approximate
$Z_{\beta\gamma}(\vec{x}+\vec{X})\approx
                       2\pi\varepsilon_{\beta\gamma}\delta(\vec{x}+\vec{X})$.
The result is $2\pi\omega$. Thus the total helicity
per unit lenght is $2\times2\pi\omega$. If we take into account
that the straight linear vortex and the helix are linked once on a distance
of $\frac{2\pi}{\omega}$ we conclude that a helicity per one link
amounts to $2(-2\pi)^{2}$ what is nothing else than
the expected $2\Phi^{2}$.

   We have calculated helicity for large $R$. As the helix radius $R$ is
turned to zero the helicity per unit lenght also smoothly diminishes to zero.
For $R=0$ we have just a single vortex with a winding number 2. Now the vortex
can be split into two unit vortices which can be moved apart. In this way
we have constructed a family of solutions which interpolates between a
linked pair of vortices and straight vortices standing far apart. The helicity
changed continously from its initial value to zero. We have considered
only infinitely long vortices but the result should be qualitatively the same
for very large vortex loops.
Two loops which were initially linked can be delinked into two fairly separated
loops with a continuous change in helicity. For this way of delinking
one would require much more restrictive initial conditions than for
intercommutation of loop segments so we think intercommutation
to be the dominant process.

    Let us summarise the scenario of baryon number violation which has been
clarified in this paper. The initial configuration is a pair of linked
$Z$-string loops. Linking of the loops enforces them to be
twisted. There are two channels of decay. The first is a decay of the linked
loops into twisted string segments terminated by monopoles. Contrary
to the statement in \cite{vachaspati} this process can take place
before delinking. If the segments untwist and shrink helicity and baryon
number will be violated.
The final configuration is a set of sphalerons and antisphalerons their
number and orientation dependent on details of string breaking and
shrinking. The CS number of a sphaleron configuration is $\frac{1}{2}$ but
only in a gauge which is unitary at infinity \cite{manklink,axenides}.
Without such a gauge the helicity of a parity-odd configuration is zero.

    The other channel is through delinking of the loops by intercommutation
or just passing one through another. In both cases we get a single loop or
two separate loops which are untwisted. Delinking has changed helicity. Now
the loops can decay but this time segments are untwisted and there is no
further helicity violation. Thus both channels of decay lead to violation of
baryon number. In collisions of infinite strings intercommutation would not
change helicity but superrelativistic collisions in which strings pass one
through another would introduce local twists and violate baryon number.

   $Aknowledgement.$ I would like to thank Henryk Arod\'z for his remarks on
the first draft of this paper and Nick Manton for useful later discussions.
This work was partially supported by KBN grants.

\end{document}